\documentclass[preprint,5p,twocolumn]{elsarticle}

\usepackage{dcolumn}
\usepackage{bm}
\usepackage{longtable}
\usepackage{multirow}
\usepackage{graphicx}
\usepackage{color}
\usepackage{amssymb,amsmath,mathtools}
\usepackage{epsfig}
\def\textm{\scriptsize\textrm}
\def\r{\mathbf{r}}

\def\a{\boldsymbol{\alpha}}

\biboptions{longnamesfirst,sort}

\graphicspath{{figures/}}

\journal{Computer physics communications}

\begin{document}

\begin{frontmatter}

 \title{A wavelet-based Projector Augmented-Wave~(PAW) method:\\  reaching frozen-core all-electron precision with a systematic, adaptive and localized wavelet basis set}

\author[paris,lbl]{T. Rangel}
\ead{trangel@lbl.gov}
\author[gren]{D. Caliste}
\author[gren]{L. Genovese}
\author[paris]{M. Torrent}

\address[paris]{CEA, DAM, DIF, F-91297 Arpajon, France}
\address[lbl]{Current affiliation: Molecular Foundry, Lawrence Berkeley National Laboratory, Berkeley, California 94720,USA}
\address[gren]{Univ. Grenoble Alpes, CEA, INAC-MEM, L\_Sim, F-38000 Grenoble, France}

\address{}

\begin{abstract}
We present a Projector Augmented-Wave~(PAW) method based on a wavelet basis set.
We implemented our wavelet-PAW method as a PAW library in the \textsc{ABINIT} package~[http://www.abinit.org] and into BigDFT~[http://www.bigdft.org].
We test our implementation in prototypical systems to illustrate the potential usage of our code.
By using the wavelet-PAW method, we can simulate charged and special boundary condition systems with frozen-core all-electron precision.
Furthermore, our work paves the way to    large-scale and potentially order-$N$ simulations within a PAW method.
\end{abstract}



\end{frontmatter}


\section{Introduction}
\label{intro}

Density Functional Theory (DFT) has recently gained popularity
due to its inherent efficiency and simplicity.
It is one of the most reliable first-principles methods to 
simulate material properties.
Presently, DFT is widely used in
the physics, chemistry and biology communities.

DFT codes have largely been developed 
to harness recent advances in modern computing.
Parallel supercomputers consisting of 
thousands of processors have become the norm.
Hence, highly-parallel architectures and efficient methods 
are desired in a DFT code.
The \textsc{BigDFT} library~\cite{genovese.2008}, which is integrated in the \textsc{ABINIT} software suite~\cite{abinit.2009},  has been conceived and implemented for massively-parallel environments in view of these large-scale calculations.

The quintessential characteristic of \textsc{BigDFT} is the use of a Daubechies~\cite{daubechies.1992}
wavelet~(WVL) basis set to 
express the Kohn-Sham~(KS) orbitals.
WVLs form a localized, systematic, orthogonal and adaptive basis set, and hence combine most of the advantages of computational basis sets for DFT.
A WVL basis set consists of a set of inter-related scaling functions and WVLs in a uniform grid of spacing $h$.
Convergence is, thus, achieved by decreasing the value
of $h$ up to the required accuracy. 
Moreover, in \textsc{BigDFT} a two-level grid is defined.
There is a fine resolution grid close to the atoms to represent the
chemical bonds and atomic orbitals, 
and secondly a coarse grid encompasses a larger volume further away from the atoms until the wavefunction vanishes.
This high-degree of adaptivity makes the BigDFT library optimal for calculations of polarized systems or systems with non Born-von Karman boundary conditions, such as material surfaces and isolated molecules.

Over the last few decades, the Projector Augmented-Wave~(PAW) method~\cite{blochl_projector_1994} has proven successful in electronic structure calculations due to its frozen-core all-electron~(AE) accuracy at low computational cost due to ultra-soft pseudopotentials~(USPP).
Within PAW, the AE wavefunction $\Psi$ is obtained from an auxiliary smooth wavefunction $\tilde{\Psi}$ by a linear transformation $\mathcal{T}$,
as $|\Psi \rangle =\mathcal{T} | \tilde{ \Psi} \rangle$.
Close to the atoms in the so called ``augmentation regions'' wavefunctions tend to have rapid oscillations. 
This can be problematic since many basis elements are usually required to express the wavefunctions within norm-conserving pseudopotentials~(NCPP).
However, within PAW a minimal basis set of atomic partial waves is sufficient to express the wavefunctions in these regions.
Due to its accuracy and high computing performance, the PAW method has gained enormous popularity and it has been implemented in several DFT codes, which use either plane-wave~\cite{abinit.2009,cp-paw,pwscf,kresse_ultrasoft_1999,socorro,tackett_projector_2001} or real-space~\cite{kromen_projector_2001,mortensen_real-space_2005,skylaris_introducing_2005,enkovaara_electronic_2010} approaches.
To our knowledge the PAW method has never been implemented within WVLs.

In this work, we detail a WFL-based PAW implementation in the \textsc{ABINIT} package using the \textsc{BigDFT} library.
By making a PAW library the method has been ported into \textsc{BigDFT} in stand-alone mode, and therefore it can be combined with different basis sets in other codes.
Our approach presents the PAW AE accuracy, exploits the wavelet adaptability and can potentially benefit from the BigDFT order-$N$\cite{mohr_daubechies_2014,mohr_accurate_2015} capabilities.

The paper is organized as follows.
We give an overview of the theory of WVL-based DFT and PAW in Section~\ref{sec.background}.
In Section~\ref{sec.methodology}, we explain the technical details of the implementation, detailing the approximations used to solve the KS equations.
Input variables used in the code are introduced to 
help readers using the code.
In Section~\ref{sec.application}, 
we show numerical tests illustrating the applicability of our code.
Finally, we draw the conclusions  and discuss future directions of this work in Section~\ref{sec.conclusions}.

\section{Background theory}
\label{sec.background}

\subsection{Wavelet-based Density Functional Theory}
In this Section, the theoretical background of the WVL formalism
is briefly presented following 
the specific choices made by the \textsc{BigDFT} library.
We emphasize that we neither discuss the advantages of wavelets as a basis set nor the main operations related to the BigDFT code, as these have been previously detailed in Refs.~\citenum{beylkin_representation_1992, neelov_efficient_2006,genovese.2008, deutsch_wavelets_2011} and \citenum{Cerioni2013}.
In this work, we rather present our implementation of PAW associated with a wavelet basis set.
\subsubsection{Basis set}

\begin{figure}[h!]
\includegraphics{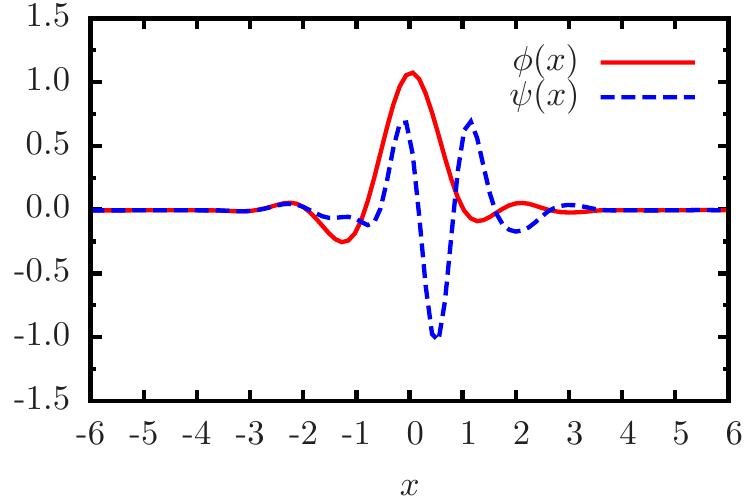}
\caption{(Color online)
Daubechies wavelet~$\psi(x)$ and scaling function~$\phi(x)$ of order 16.
}
\label{fig:wvl}
\end{figure}

We use the so called maximally symmetric Daubechies WVL basis of order 16,~\cite{daubechies.1992} since they present virtually ideal
properties of a basis set:
they are orthogonal, systematic, can represent exactly up to 8-th order polynomials, and with translational invariance.~\cite{daubechies.1992,beylkin_representation_1992,neelov_efficient_2006,genovese.2008,deutsch_wavelets_2011,Cerioni2013,genovese_multipole-preserving_2015}
In general, a WVL basis set consists of two objects: WVLs $\psi(x)$
and scaling functions $\phi(x)$.
An illustration of the WVLs and scaling function is shown in Figure~\ref{fig:wvl}.

In the present implementation we use a two-level adaptive grid.
Away from the atoms, we use a ``coarse'' basis set, which consists of three-dimensional~(3D) scaling functions on a grid with uniform mesh $h$:
\begin{equation}
 \phi_{i,j,k}(\mathbf{r})
= \phi(x/h-i)
\phi(y/h-j)
\phi(z/h-k)
\label{3d.scaling}
\end{equation} 
Close to the atoms we use a fine resolution degrees of freedom, with both scaling functions and wavelets, as high-resolution is needed to express the
chemical bonds and the atomic orbitals.
The basis set is here
augmented by a set of seven 3D WVLs:
\begin{eqnarray}
 \psi^1_{i,j,k}(\mathbf{r}) &=& \psi(x/h-i) \phi(y/h-j) \phi(z/h-k), \nonumber\\
 \psi^2_{i,j,k}(\mathbf{r}) &=& \phi(x/h-i) \psi(y/h-j) \phi(z/h-k), \nonumber\\
 \psi^3_{i,j,k}(\mathbf{r}) &=& \psi(x/h-i) \psi(y/h-j) \phi(z/h-k), \nonumber\\
 \psi^4_{i,j,k}(\mathbf{r}) &=& \phi(x/h-i) \phi(y/h-j) \psi(z/h-k), \nonumber\\
 \psi^5_{i,j,k}(\mathbf{r}) &=& \psi(x/h-i) \phi(y/h-j) \psi(z/h-k), \nonumber\\
 \psi^6_{i,j,k}(\mathbf{r}) &=& \phi(x/h-i) \psi(y/h-j) \psi(z/h-k), \, \mbox{and}\nonumber\\
 \psi^7_{i,j,k}(\mathbf{r}) &=& \psi(x/h-i) \psi(y/h-j) \psi(z/h-k).
\end{eqnarray}

\begin{figure}[h!]
\includegraphics[width=\columnwidth]{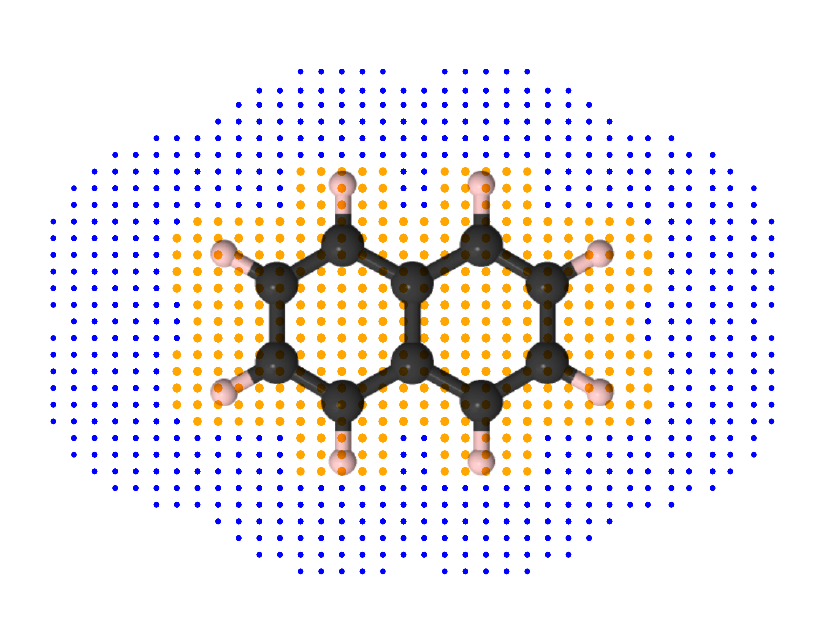}
\caption{(Color online)
Adaptive grid around a naphthalene molecule.
The high- and low- resolution grids points
are shown in orange and blue dots, respectively.
Carbon and Hydrogen 
atoms are drawn with black and pink
spheres, respectively.
}
\label{fig:adaptive}
\end{figure}

Due to the exponential localization of wavefunctions, 
far from the atoms, where the wavefunctions vanish, no basis elements are used.
To illustrate, in Figure~\ref{fig:adaptive} the 2-level adaptive grid around a naphthalene molecule is shown.

The wavefunctions $\Psi( \mathbf{r} )$ are thus 
expanded as:
\begin{eqnarray}
\Psi(\mathbf{r})=
\sum_{i_1,i_2,i_3}
s_{i_1,i_2,i_3}\phi_{i_1,i_2,i_3}(\mathbf{r}) + \nonumber\\
\sum_{j_1,j_2,j_3}
\sum_{\nu=1}^7
d^\nu_{j_1,j_2,j_3}
\psi^\nu_{j_1,j_2,j_3}(\mathbf{r}),
\label{wvl-expansion}
\end{eqnarray}
where $s_{i_1,i_2,i_3}$ and $d^\nu_{j_1,j_2,j_3}$
are expansion coefficients.
The sum over $i_1$, $i_2$ and $i_3$
runs over all points on the coarse grid
and the sum over $j_1$, $j_2$ and $j_3$ 
runs over points on the fine grid.

In our implementation we exploit the {\it separability} property of the basis functions, in which the 3D scaling functions basis set, being a product decomposition of one-dimensional~(1D) scaling functions/wavelets, is separable in the three Cartesian directions, see Eq.~(\ref{3d.scaling}).
\textsc{BigDFT} treats efficiently {\it Gaussian pseudopotentials} of the Goedecker-Teter-Hutter~(GTH)~\cite{goedecker_separable_1996}
and Hartwigsen-Goedecker-Hutter~(HGH)~\cite{hartwigsen_relativistic_1998} kinds, since the intrinsic separability of both the basis set and Gaussian pseudopotentials allows for the simplification of several 3D operations into a sum of 1D products.
Following the same spirit, for PAW it makes sense to use non-local projectors expressed as sum of Gaussians, as described in Section.~\ref{sec.non-local}.

\subsection{The PAW method}
In this Section, we briefly review the PAW formalism.
We adopt the notation of Ref.~\citenum{torrent_implementation_2008}.

\subsubsection{The PAW transformation}
In the PAW scheme,
the true AE wavefunctions $\Psi$ 
are obtained from the auxiliary wavefunctions $\tilde{\Psi}$
(known as pseudo-wavefunctions)
by applying a linear operator $\mathcal{T}$, expressed as a sum of atom-dependent contributions:
\begin{eqnarray}
 |\Psi \rangle &=& \mathcal{T} | \tilde{\Psi} \rangle =
 \left( 1+\sum_R \mathcal{T}_R  \right) | \tilde{\Psi} \rangle,
\label{eq.transformation}
\end{eqnarray}
where $\mathcal{T}_R$ are local contributions acting only
in an {\it augmentation region} around the atom.
The PAW transformation takes the form:
\begin{equation}
 |\Psi\rangle = 
 |\tilde{\Psi} \rangle
 + \sum_i
 \left( |\phi_i \rangle -  | \tilde{ \phi_i} \rangle \right)
 \langle \tilde{p}_i | \tilde{\Psi} \rangle,
\label{eq:psi-paw}
\end{equation}
where the $i$ index runs on atom position $\mathbf{R}$, 
angular momentum ($l$, $m$)
and additional index $m'$ for different partial waves with 
the same angular momentum and atom site.
The AE $\phi_i$ and pseudo $\tilde{\phi_i}$ partial waves
are identical outside the augmentation region.
The partial waves and the projectors $\tilde{p_i}$ 
are calculated in a spherical grid inside 
the augmentation region
separating the radial and angular parts:
\begin{eqnarray}
\hspace{-0.1in}
 \phi_i(\mathbf{r}) = 
\frac{\phi_{n_i,l_i} (r)}{r}
S_{l_i,m_i}; \hspace{0.1in}
\tilde{\phi}_i(\mathbf{r}) = 
\frac{\tilde{\phi}_{n_i,l_i} (r)}{r}
S_{l_i,m_i} (\hat{r})
\end{eqnarray}
\begin{equation}
\hspace{-0.1in}
\tilde{p}_i(\mathbf{r}) =
\frac{\tilde{p}_{n_i,l_i} (r)}{r}
S_{l_i,m_i}(\hat{r}), \hfill 
\end{equation}
where $S_{l,m}(\hat{r})$ are the real spherical harmonics.
A more extensive review of the method can be found in 
the original paper of Bl\"ochl~\cite{blochl_projector_1994} and 
the implementation in ABINIT within plane-waves
is detailed in Refs.~\cite{torrent_implementation_2008,Torrent20101862}.

From Eq.~(\ref{eq:psi-paw}), the AE valence charge density becomes:
\begin{equation}
 n_v=\sum_{nk} f_{nk} |\Psi_{nk}|^2 =
 \tilde{n} + n^1 - \tilde{n}^1,
 \label{eq:nv}
\end{equation}
where $f_{nk}$ is the occupation number of 
band $n$ at $k$-point $k$.
The pseudized density $\tilde{n}$ is simply 
$\tilde{n}=\sum f_{nk} |\tilde{\Psi}_{nk}|^2 $, 
akin the NCPP charge density.
$n^1$ and $\tilde{n}^1$ are the AE and pseudized on-site densities, respectively.
These are only defined in the augmentation regions.
Note that the superscript "$1$" refers to atomic quantities.

To evaluate the on-site densities, the occupancy matrix $\rho_{ij}$ is required~\cite{torrent_implementation_2008},
\begin{equation}
\rho_{ij}= \sum_{nk} f_{nk}
(c^{i}_{nk})^*
c^{i}_{nk} ;
\,\,\,
c^{i}_{nk}= \langle \tilde{p}_i | 
\tilde{\Psi}_{nk}
\rangle.
\end{equation}
The {\it compensation charge} $\hat{n}$~\cite{kresse_ultrasoft_1999,blochl_projector_1994} is added to the pseudized densities $\tilde{n}$ and $\tilde{n}^1$.
The purpose of $\hat{n}$, as introduced in Ref.~\cite{kresse_efficient_1996}, is to add the correct amount of charge moments to the valence pseudo density $\tilde{n}$ so that outside the augmentation region of all the atoms, the Coulomb (or Hartree) potential for the sum of the valence pseudo and compensation charge densities $(\tilde{n}(\mathbf{r})+\hat{n}(\mathbf{r}))$ is the same as that for the fully nodal valence electron density $n(\mathbf{r})$.
{\it i.e.,}  $\hat{n}$ is constructed ensuring that the Coulomb potential created from the on-site densities cancels out outside the augmentation regions, avoiding electrostatic interactions between PAW spheres.
Further, a high resolution around the atoms is needed to ensure that the pseudo densities inside and outside the augmentation regions exactly cancel out.
Therefore, a {\it double-grid} technique~\cite{kresse_from_1999} is usually used to represent the density and potential terms.

\subsubsection{The PAW Kohn-Sham Hamiltonian}
\label{sec.ks}
The Kohn-Sham~(KS) Hamiltonian $\mathcal{H}$ is defined as~\cite{kresse_efficient_1996,vanderbilt_soft_1990}:
\begin{equation}
\mathcal{H}= T+ V_{\textm{L}}+V_{\textm{NL}}
\end{equation}
where the effective potential 
consists of the kinetic~$T$, the local~(L)
and the non-local~(NL) potentials.
The local potential $V_{\textm{L}}$ is given by
\begin{equation}
 V_{\textm{L}}= V_{\textm{xc}} + V_{\textm{H}} + \sum_i^{\textm{ion}} V^{(i)}_{\textm{loc}},
\end{equation}
where $V_{\textm{xc}}$, $V_{\textm{H}}$ 
and $V^{(i)}_{\textm{loc}}$
are the exchange-correlation, Hartree and local-ionic potentials,  respectively.
The non-local potential $V_{\textm{NL}}$ is
\begin{equation}
V_{\textm{NL}}= \sum_{ij} | \tilde{p}_i \rangle D_{ij} \langle \tilde{p}_j |,
\label{eq.nl}
\end{equation}
where, 
the non-local coefficients $D_{ij}$ 
are calculated at each self-consistent field iteration.

Since the KS wavefunctions  $|\Psi_{nk}\rangle$ are non-orthogonal, 
the following generalized eigenvalue equation is solved,
\begin{equation}
 \left( \mathcal{H} - \epsilon_{nk} S \right) | \Psi_{nk} \rangle =0,
\end{equation}
where $\epsilon_{nk}$ is the eigenvalue of band $n$ at $k$-point $k$, and the overlap matrix $S$ is defined as:
\begin{equation}
S= 1+\sum_{ij}  |\tilde{p}_j \rangle s_{ij} \langle \tilde{p}_i |,
\end{equation}
with $s_{ij}= \langle \phi_i | \phi_j \rangle - \langle \tilde {\phi}_i | \tilde{\phi}_j\rangle$.

\section{Methodology}
\label{sec.methodology}

\begin{figure}
\includegraphics[width=\columnwidth]{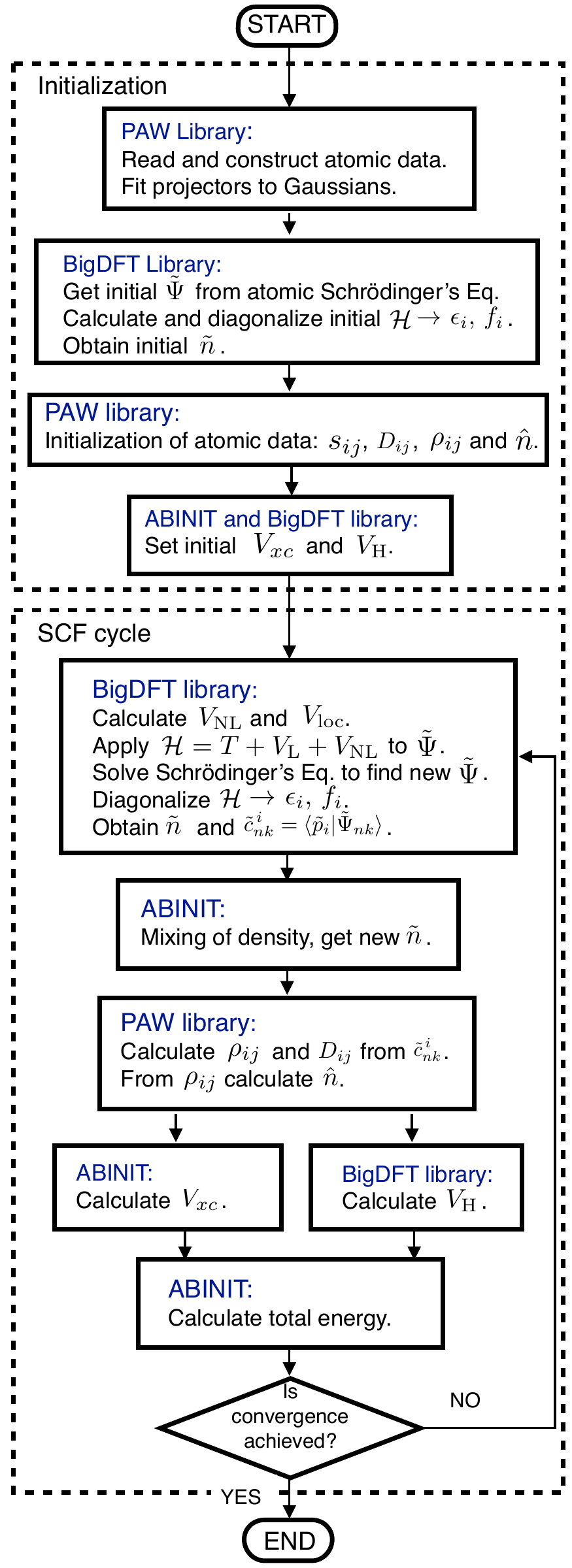}
\caption{Flow-chart illustrating the 
 the WVL-PAW method. 
The code/library used at each step in the
Initialization or the Self-Consistent Field (SCF) cycle
are indicated in blue fonts.
}
\label{fig:flowchart}
\end{figure}

In this Section, we describe the 
basic steps of our   
WVL-PAW method.
The flowchart of this
implementation is presented
in 
Figure~\ref{fig:flowchart}.

\subsection{Local ionic potential}
\label{sec.local-potential}
\begin{figure}[h!]
\includegraphics{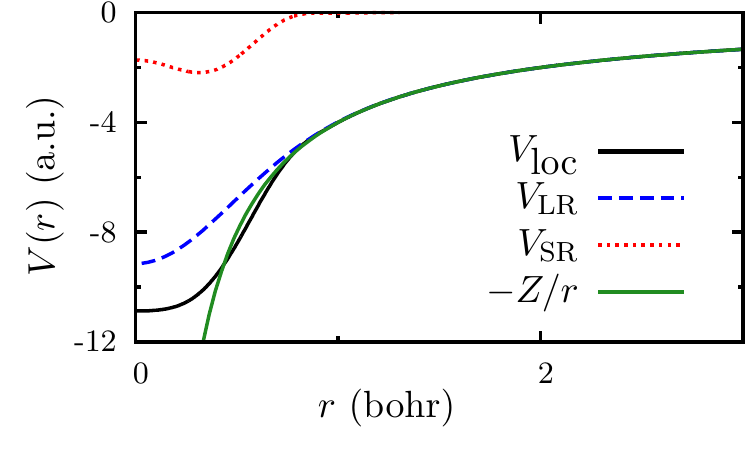}
\caption{(Color online)
Local ionic potential $V_{\textm{loc}}(r)$ 
(solid black lines) and its 
long- $V_{\textm{LR}}(r)$ (dashed blue lines)
and short-range $V_{\textm{SR}}(r)$ (dotted red lines)
components
for Carbon with $r_\ell=0.35$ bohrs.
}
\label{fig:vloc}
\end{figure}

The local ionic potential $V_{\textm{loc}}^{(i)}$ 
for each atom indexed by $i$ is calculated by
the PAW atomic-data generator,
and hence read by \textsc{ABINIT} in a radial grid.
Aiming for linear-scaling, we circumvent the explicit evaluation of the local ionic potential which presents a slow decay, as shown in Figure~\ref{fig:vloc}.
Instead, we calculate the potential locally by dividing it into 
long- (LR) and 
short-range (SR) components,
\begin{equation}
V^{(i)}_{\textm{loc}}= V^{(i)}_{\textm{SR}} + V^{(i)}_{\textm{LR}}.
\label{vloc-components}
\end{equation}
The LR part is chosen 
to be equal to the corresponding GTH-HGH pseudopotentials LR component,
\begin{eqnarray}
V^{(i)}_{\textm{LR}}(\r)=
-\frac{Z_i}{r}
\mbox{erf}\left(
\frac{r}{\sqrt{2} r_{\ell}}
\right),
\label{vlr}
\end{eqnarray}
being a smooth function with the correct decay of $-Z_i/r$, where $Z$ is the atomic number (see Figure~\ref{fig:vloc}).
The LR potential is associated to a 
given density $n_{\textm{LR}}$ via the
Poisson equation 
$\Delta V_{\textm{LR}} = -4\pi n_{\textm{LR}}$, where
\begin{equation}
  n^{(i)}_{\textm{LR}}(\r)
=
-\frac{1}{(2\pi)^{3/2}}
\frac{Z_i}{r^3_\ell}
e^{-\frac{r^2}{2r^2_\ell}},
\label{rho-loc}
\end{equation}
and the radius $r_\ell$ is chosen to be between $0.2$ and $1.0$ bohrs.
Since the LR density is localized around the atoms, we can calculate locally the corresponding LR potential in a two step procedure;
We first calculate $n_{\textm{LR}}$ at each atom, and afterwards, we find the corresponding potential in the entire the simulation box via the \textsc{BigDFT} Poisson-solver~\cite{Cerioni2013}.
Moreover, the smooth shape of Eq.~(\ref{rho-loc}) is particularly convenient to achieve a  fast convergence in the evaluation of $V_{\textm{loc}}$ with respect to the grid-spacing $h$.

From Eq.~(\ref{vloc-components}),
the SR potential is given by
\begin{equation}
  V^{(i)}_{\textm{SR}}= V^{(i)}_{\textm{loc}}-V^{(i)}_{\textm{LR}}, 
\end{equation}
and can be easily evaluated locally at each atom.
The LR part is analytic, see Eq.~(\ref{vlr}) and we calculate  $V^{(i)}_{\textm{loc}}$ using splines.
To avoid the divergence, the term $V^{(i)}_{\mbox{\tiny{LR}}}(\r=0)$ is evaluated using a quadratic interpolation of Eq.~(\ref{vlr}) from 3 points close to the origin.
In summary, by dividing $V_{\textm{loc}}$ into LR and SR components, which can be evaluated locally at each atom, the calculation of $V_{\textm{loc}}$ scales linearly with the number of atoms $N_{\textm{at}}$.

\subsection{Non-local potential}
\label{sec.non-local}

We calculate the NL potential in Eq.~(\ref{eq.nl}),
\begin{eqnarray}
V_{\textm{NL}}= 
\sum_{ij} | \tilde{p}_i \rangle D_{ij} \langle \tilde{p}_j |, \nonumber
\end{eqnarray}
by generalizing the method of \textsc{BigDFT} to complex-Gaussian PAW pseudopotentials.
The application of the Hamiltonian ($\mathcal{H} |\Psi \rangle$)
requires the calculation 
of wavefunction projections $\tilde{c}^i_{nk}=\langle \tilde{p}_i | \Psi_{nk} \rangle$.
Both wavefunctions and projectors
are expanded on the WVL basis in order to do this operation by simple scalar products, where the expansion coefficients for projectors are: ~\cite{goedecker_wavelets_1998}
\begin{equation}
 \int \tilde{p}(\mathbf{r}) \phi_{i_1,i_2,i_3}(\mathbf{r})\mbox{d}\mathbf{r}
 \hspace{0.1in}
 \mbox{and} 
 \hspace{0.1in}
 \int \tilde{p}(\mathbf{r}) \psi^{\nu}_{i_1,i_2,i_3}(\mathbf{r})\mbox{d}\mathbf{r}.
\label{eq.integral}
\end{equation}

In \textsc{BigDFT},
the integrals in Eq.~(\ref{eq.integral}) are simplified 
by using pseudopotentials with projectors of the
form:
\begin{equation} 
 \tilde{p}(\mathbf{r})=e^{-cr^2} x^{\ell_x}y^{\ell_y}z^{\ell_z};
 \hspace{0.2in} c \in \mathbb{R}.
\end{equation}
Following the same spirit,
 PAW projectors are {\it fitted} to the general form:
\begin{equation} 
 \tilde{p}(\mathbf{r})
\approx
x^{\ell_x}y^{\ell_y}z^{\ell_z} \sum_j^{N_g} a_j e^{b_jr^2} ;
 \hspace{0.2in} a_j,b_j \in \mathbb{C},
\label{eq.projectors-gauss}
\end{equation}
where the number of complex Gaussian functions $N_g$ is to be determined.
This analytical form simplifies the 3D integrals
in Eq.~(\ref{eq.integral})
into a sum of products of three 1D integrals.
\begin{eqnarray}
 \int \tilde{p}(\mathbf{r}) 
\phi_{i_1,i_2,i_3}(\mathbf{r})\mbox{d}\mathbf{r} &=& 
\sum_j^{N_g}
W_{i_1}\left(a_j,b_j,\ell_x\right) \times \nonumber\\
&&\hspace{-0.6in}
W_{i_2}\left(a_j,b_j,\ell_y\right) \times
W_{i_3}\left(a_j,b_j,\ell_z\right), \\
W_{i_k}\left(a,b,\ell\right) &=&
\int_{-\infty}^{+\infty} a e^{b t^2} t^\ell 
\phi(t/h-k) \mbox{d}t. \nonumber\\
\end{eqnarray}
The 1D integrals are calculated following the \textsc{BigDFT} scheme,
which is accurate to machine precision.
Moreover, the fitting procedure introduced above is robust enough to obtain the desired accuracy in total energies, as explained more thoroughly in Section~\ref{sec.proj-gauss}.

\subsection{Direct minimization method} 
\label{diis}

In a direct minimization approach, the gradient $|g_i\rangle$ of the total energy with respect
to the $i$th wavefunction $|\Psi_i\rangle$ is defined 
in terms of the Hamiltonian $\mathcal{H}$ and the overlap $S$
operators:
\begin{eqnarray}
 |g_i\rangle &=&
\mathcal{H} | \tilde{\Psi}_i \rangle - 
\sum_j \Lambda_{ij} S |\Psi_j \rangle, \label{gradient} \\
\Lambda_{ij}&=&
\langle \Psi_j | \mathcal{H} |\Psi_i \rangle
\end{eqnarray}
In \textsc{BigDFT}, the preconditioned gradient $|\tilde{g}\rangle$,
is found by solving the equation
\begin{equation}
\left(  \frac{1}{2} \nabla^2  - \epsilon_i \right)
\tilde{g_i}(\mathbf{r})=g_i(\mathbf{r})\;.
\end{equation}
The information coming from the wavefunction gradient
is usually combined in the context of the 
{\it direct inversion in the iterative subspace~(DIIS)}, method~\cite{peter_convergence_1980,kresse_efficient_1996}
where the wavefunction $i$ at iteration $M+1$ is found by
suitable linear combination of the previous $M$ trial functions 
$|\Psi^j_i\rangle$ and $|\tilde{g}^j\rangle$.:
\begin{equation}
|\Psi^{M+1}_i\rangle = | \overline{\Psi}^M \rangle + 
\lambda \overline{\tilde{g}}^M \rangle,
\end{equation}
with
\begin{eqnarray}
 |\overline{\Psi}^M \rangle= \sum_{j=0}^{M}
\alpha_j |\Psi^j_i\rangle,
\hspace{0.2in}
 |\overline{\tilde{g}}^M \rangle= \sum_{j=0}^{M}
\alpha_j |\tilde{g}^j_i\rangle.
\end{eqnarray}
With the exceptions of the presence of the operator $S$ in Eq.~\eqref{gradient}, all the wavefunction optimization flowchart is identical to the Norm-Conserving approach.

\subsection{PAW atomic datasets}
\label{sec.atomic-data}
The first step in a PAW calculation is to 
read the atomic-data files (containing the description of atoms), or roughly speaking the "pseudopotentials", previously calculated.
Numerous  databases of atomic datasets are available and 
can be used within the present
WVL-PAW implementation~\cite{jollet_generation_2014,xml-gpaw,garrity_pseudopotentials_2014,pwpaw-format}.

For PAW, atomic-data files for WVLs and PWs are almost identical;
In the PW case, projectors are given in a radial grid whereas in the WVL case projectors are expressed as a sum of complex Gaussian functions.

To support all atomic-data file formats, \textsc{ABINIT} contains now a converter  from the conventional format (with projectors on a grid) to the Gaussian format.
First projectors  are fitted to a sum of Gaussian functions, as explained in ~\ref{sec.proj-gauss}.
Later, the Gaussian coefficients are written into a  new atomic-data file.
This file can be used in further calculations to avoid running the fitting procedure at every time.
From the user point of view, the above procedure is transparent.

\section{Fitting the PAW projectors}
\label{sec.proj-gauss}

As previously discussed in Sec.~\ref{sec.non-local}, PAW non-local projectors are fitted to a sum of Gaussians with complex exponents in order to simplify scalar products with wavefunctions.
Indeed, projectors expressed as linear combinations of Gaussian functions become analytical, and hence can be easily converted to WVL space.
Moreover, since both projectors and Daubechies WVLs are separable in the three Cartesian  directions, internal products $\langle \tilde{p}_j | \phi_i \rangle $ are reduced to 1D operations.

In this work, we use the Levenberg-Marquardt formalism~\cite{levenberg_method_1994} to fit the projectors to a Gaussian form.
A given set of points ($x_i,\tilde{p}_i$) are fitted to a given function f($x_i,\a$), where $\a$ is a vector of independent and/or dependent parameters.
As in other regression methods, the sum of the squares of deviations $S(\a)$ is minimized, where
\begin{equation}
 S(\a) = \sum_{i=1}^{N_i}
 \left[
 \tilde{p}_i - f(x_i,\a)
 \right]^2.
\label{chisq}
\end{equation}

The radial part of the projectors  $\tilde{p}(r)$ is fitted to the analytical expression in Eq.~(\ref{eq.projectors-gauss}):
\begin{equation} 
 \tilde{p}(r)
\approx
x^{\ell_x}y^{\ell_y}z^{\ell_z} \sum_j^{N_g} a_j e^{b_jr^2} ;
 \hspace{0.2in} a_j,b_j \in \mathbb{C}. \nonumber
\end{equation}
As other minimization techniques,
the fitting algorithm is quite 
sensitive to the initial-guess.
Moreover,  a function which decays to zero (such as PAW projectors) is generally difficult to fit to periodic functions (sinus and cosinus functions).
Therefore, we use a real Gaussian function as an envelope of a sum of sinus and cosinus functions, as follows,
\begin{align}
 \tilde{p}(r)
\approx &
\sum_j^{N_g}
a_{1,j} e^{-a_{2,j} x^2}
\left(
a_{3,j} sin\left( k_j x^2 \right)
a_{4,j} cos\left( k_j x^2 \right)
\right) ;&
\nonumber\\
&
a_1,a_2,a_3,a_4 \in \mathbb{R} &
\label{eq:fit}
\end{align}
To obtain a small $S$,
the envelope function is
constrained to decay almost to zero at the PAW radius.
Further, the $k^j$ are fixed to an exponential series ({\it i.e.,} $k=1.1^j$) to enhance the sinus \& cosinus basis completeness, and hence avoiding duplicate basis set elements.

\begin{figure}[h!]
\includegraphics{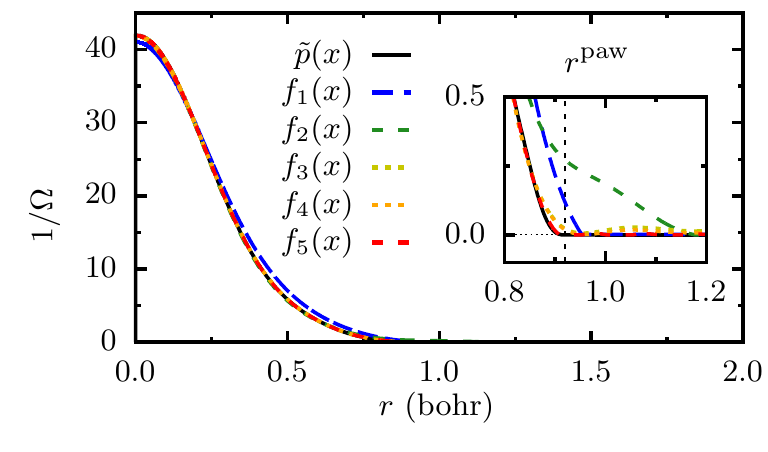}
\includegraphics{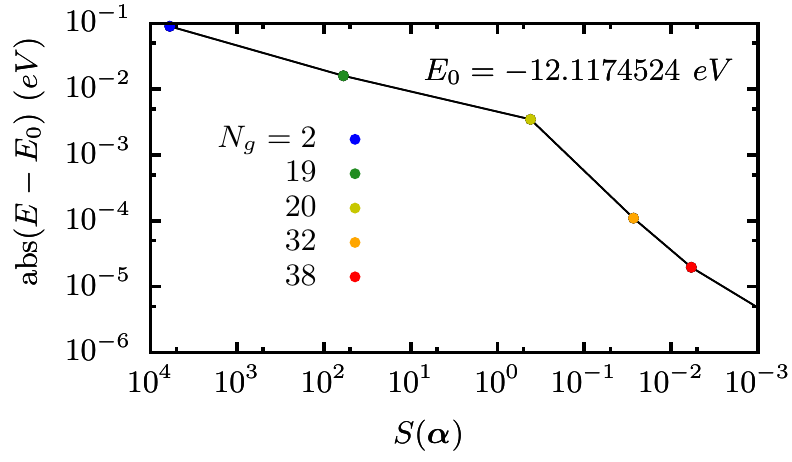}
\caption{(Color online)
The $1s$ Hydrogen NL PAW projector~[$\tilde{p}(x)$]
is fitted to a sum of complex Gaussians.
The number of terms $N_g$ in Eq.~(\ref{eq:fit}) 
is increased to achieve a better fit [a  smaller $S(\a)$].
{\bf Top:} 
The original function [$\tilde{p}(x)$] is shown
in solid-black lines on a radial grid.
The fitted functions using $N_g$ equal to
2  [$f_1(x)$] [dashed-blue line],
19 [$f_2(x)$] [dashed-green line],
20 [$f_3(x)$] [dashed-yellow line],
32 [$f_4(x)$] [dashed-orange line] and
38 [$f_5(x)$] [dashed-red line]
are also shown.
In the inset, the area close 
to the PAW radius, $r^{\textm{paw}}$ is zoomed-in.
{\bf Bottom:}
Error in total energies (in hartrees) with respect to a plane-waves-PAW calculation for Hydrogen in a box.
The results for several $N_g$ values corresponding 
to different $S(\a)$ are shown.
Increasing the number of Gaussian functions improves the 
accuracy in the calculated total energy,
as expected.
}
\label{fig:fitting}
\end{figure}

The fitting procedure is 
overall satisfactory.
As expected, the accuracy
can be improved by increasing the
number of complex Gaussian functions $N_g$.
For instance, in Figure~\ref{fig:fitting}
the $1s$ NL-projector for Hydrogen
is fitted to a sum of complex Gaussians. For simplicity,  spin-polarization is not taken into account.
As expected, for larger $N_g$  smaller $S(\a)$ may be obtained, hence, the error due to fitting of $\tilde{p}$ in total energy calculations decreases, reaching the same total energy, up to machine-precision, than the one obtained with a plane-waves-PAW calculation.
In particular, the number of Gaussian functions $N_g$ is a convergence parameter in the simulation, as exemplified in the next Section.

\section{Numerical tests}
\label{sec.application}

\begin{figure}[h!]
\begin{minipage}[c]{0.25\textwidth}
\includegraphics[width=\textwidth]{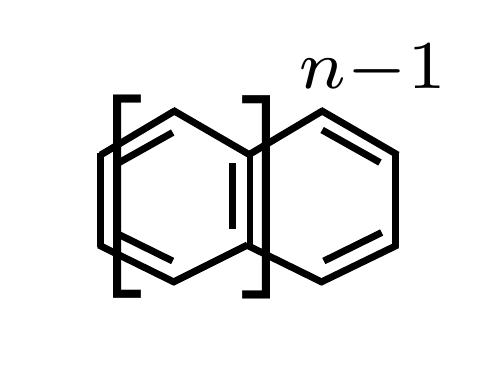}
\end{minipage}\hspace{0.1in}
\begin{minipage}[c]{0.15\textwidth}
\caption{The acenes general formula.}
\label{fig:acenes}
\end{minipage}
\end{figure}

In this Section, we show tests on simple systems to elucidate the performance of our implementation and the effect of the Gaussian expansion of the PAW non-local projectors.
In particular, we present calculations of total-energy differences in finite systems, since wavelets are particularly adapted for charged species in non-periodic boundary conditions.

We first show that our method provides results in agreement with previous and well-established DFT codes, for the sake of validation.
With this aim, we compare total-energy differences calculated with our WVL-PAW implementation and with HGH PPs with WVL basis.
Here, we study the acene family of molecules: from benzene to hexacene, as a prototype finite system.
The general formula is shown in Figure~\ref{fig:acenes}.
In Table~\ref{table:acenes}, we show the  ionization potential~(IP) of the acenes calculated as
\begin{equation}
\mbox{IP}
= \frac{1}{2} \left( E(N+2) - E(N) \right),
\label{eq:ip}
\end{equation}
and neglecting spin as a first-order approximation.
The IPs calculated with our WVL-PAW implementation are within 0.05~eV of those calculated with HGH PPs.
As a further check, the resulting IPs were verified with the former PW-PAW  implementation in \textsc{ABINIT}. 

\begin{table}
\begin{tabular}{lcc}
\hline\hline\\
Molecule& \multicolumn{2}{c}{Ionization potential (eV)} \\
			&PAW & HGH\\
Benzene  	& 13.05 & 12.99\\
Naphthalene & 10.87 & 10.81\\
Anthracene	&  9.55 &  9.57\\
Tetracene	&  8.76 &  8.77\\
\hline\hline
\end{tabular}
\caption{Ionization potential~(IP) of the oligoacenes calculated with energy differences of the neutral and charged species within the LDA.
Several methods are considered:
our WVL-PAW implementation and HGH PPs with WVL basis, for comparison.}
\label{table:acenes}
\end{table}

\begin{figure}[h!]
\includegraphics[width=0.5\textwidth]{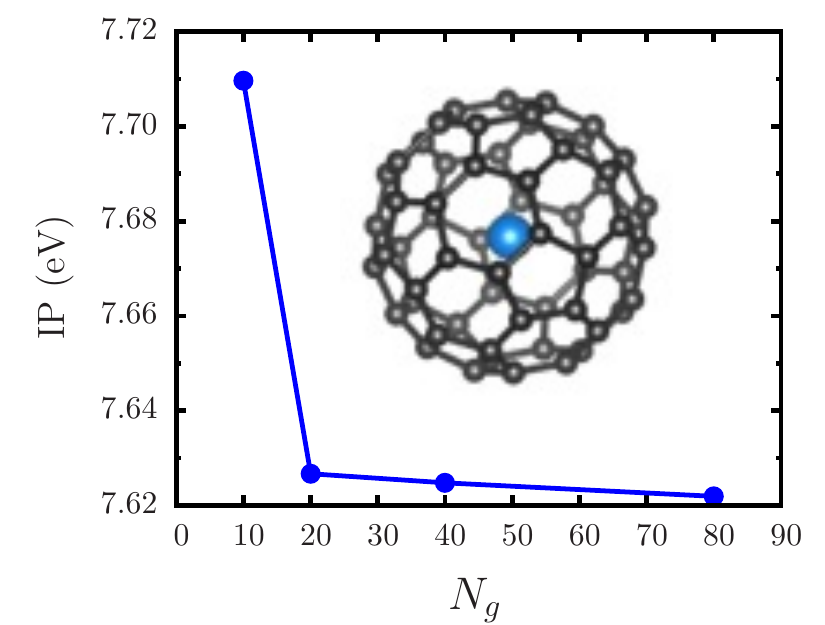}
\caption{Endohedral doped fullerene with a Ti atom enclosed at its center; carbon and Ti atoms are shown in gray and blue spheres, respectively. We show the calculated ionization potential~(IP) energy with respect to the number of complex Gaussian projectors ($N_g$ terms).}
\label{fig:c60-ti-ng}
\end{figure}

Having verified the accuracy of our method, we now  simulate a relatively more complex and inhomogeneous case consisting of a C60 molecule doped with a Ti atom at its center. 
In Figure~\ref{fig:c60-ti-ng}, we show the convergence of the calculated ionization potential energy~(IP) with respect to the number of complex Gaussian projectors.
As before, the IP is calculated from total energy differences with Eq.~\ref{eq:ip}.
Interestingly, a relatively small number of $N_g=10$ terms is needed to converge the IP energy to 0.1~eV. This proves that the Gaussian fit of the PAW projectors might provide an interesting strategy to express in analytic form a $\mathcal{T}_R$ operator.
Work is in progress to identify the better strategy to reduce the computational overhead in the case when the fit needs a large (e.g. bigger than 40) number of Gaussians.

In summary, we showed that our WVL-PAW implementation is accurate and efficient and is appropriate to simulating complex in-homogeneous systems in non-periodic boundary conditions.

\section{The PAW library}
While implementing the WVL-PAW code, 
the core PAW routines of 
\textsc{ABINIT} 
were assembled, modularized,  and disentangled from the PWs part of the code. 
The outcome of this work is a basis-independent, modular and and stand-alone {\it PAW library} written in Fortran 2003.
It is worth noticing that the 
XC potential is  calculated by using 
\textsc{ABINIT} or the \textsc{Libxc} package~\cite{marques_libxc_2012}.
The library is now stable and ready to be used in other codes. 

We emphasize that the PAW library is basis independent, and hence it can be easily ported to another basis, as proven by our implementation of PAW in BigDFT. 
The only part that remains dependent on the given basis set are the projections of the pseudo-wavefunctions onto the non-local projectors $\langle \tilde{\Psi}| \tilde{p}_i\rangle$.
From the knowledge of these projections, all quantities involved in the PAW formalism can be computed: occupancy matrix, on-site densities, self-consistent contribution to the non-local operator, etc. 
And this is exactly the purpose of the PAW library.
For more details, please consult the \textsc{ABINIT} documentation, or the code sources, all modules are self-explanatory.

In the next, the basic modules of the PAW library are briefly described.

Modules related to basic datatypes:
\begin{itemize}
\item {\tt m\_pawrad}:
contains all functions related to the PAW radial meshes, such as datatypes and derivation/integration routines for the different kinds of meshes (linear or logarithmic).
\item {\tt m\_pawtab}:
used to define tabulated PAW data which is read from atomic data files, such as core- and local- potentials and PAW projectors, etc.
\item {\tt m\_pawang}:
contains definitions and functions related to the angular mesh discretization of the PAW spheres.
\item {\tt m\_paw\_ij}:
contains objects expressed in the on-site partial-wave basis, {\it e.g.,} the calculated $D_{ij}$ terms.
\item {\tt m\_paw\_an}:
 on-site potentials and densities are defined here. 
The potentials/densities are stored in terms of  angular mesh or angular moments for each atom.
\end{itemize}

Modules related to high-level objects:
\begin{itemize}
\item {\tt m\_pawpsp}:
module to read PAW atomic data files.
Used to read and calculate data inside the augmentation spheres, such as potentials and its derivatives.
\item {\tt m\_pawcprj}:
calculates, stores and manipulates $\tilde{c}^i_{nk}$.
Here we include routines to compute the on-site contribution to the overlap between two states and MPI communication routines dealing with $\tilde{c}^i_{nk}$ objects.
\item {\tt m\_pawrhoij}:
computes and symmetrizes occupancy matrix 
$\rho_{ij}=\sum_{nk} 
f_{nk} (\tilde{c}^i_{nk})^* 
\tilde{c}^j_{nk}$.
This module also contains MPI routines to send/distribute/gather the $\rho_{ij}$ occupations. 
\item {\tt m\_pawdij}:
calculates all  contributions to the
$D_ij$ non-local pseudopotential terms.
This module also computes local Hubbard-U and local exact-exchange contributions to the PAW onsite-potentials.
\end{itemize}

Low-level modules:
\begin{itemize}
\item {\tt m\_paw\_gaussfit}:
contains routines used to fit the numeric NL projectors to sums of complex Gaussians, see~\ref{sec.proj-gauss}.
\item {\tt m\_pawxc}:
computes the exchange-correlation
potential/energy in
the augmentation regions using
developments over spherical harmonics,
see Ref.~\cite{torrent_implementation_2008}.
\item {\tt m\_paw\_finegrid}:
contains all routines performing
operations ({\it e.g.,} integrations) on the fine-grid around each atom.
\end{itemize}

\section{Conclusions}
\label{sec.conclusions}

In this work we presented a PAW method in a WVL basis set.
In order to take advantage of WVL properties, we modified the PAW non-local operator to a sum of Gaussians. 
Hence, the application of the non-local part of the Hamiltonian is performed analytically, reducing  computational costs.
Our WVL-PAW method was implemented in \textsc{ABINIT} using the \textsc{BigDFT} library and by creating a PAW library the method was ported into \textsc{BigDFT} in stand-alone mode.
In addition, the PAW library is stable and can be used in other codes.
In summary, our new method presents the PAW frozen-core AE accuracy and the WVL adaptability, locality and systematic convergence.
This opens up the possibility to treat
large heterogeneous systems and large molecules within PAW and of potentially having
an order-$N$ code in PAW.

\section{Acknowledgments}
T. Rangel thanks Jean-Michel Beuken for his technical support on \textsc{ABINIT} and Alessandro Mirone for his help and discussions on the Gaussian fitting procedure.
This work was supported by the ANR NEWCASTLE project, grant ANR- 2010-COSI-005-01  of the French National Research Agency. 
\appendix

\section{The BigDFT Poisson solver 
for 
 planewave based-PAW calculations}
\label{sec.poisson}

Finite, isolated or low-dimensional systems are often simulated using supercells in periodic boundary conditions~(BC), such as in plane-wave approaches.
In order to eliminate spurious electrostatic interactions due to neighboring cells, numerous methods are well established and have been documented in Refs.~\cite{leslie_energy_1985,makov.1995,jarvis_supercell_1997,kantorovich_elimination_1999,bengtsson_dipole_1999,martyna_reciprocal_1999,schultz_charged_2000,nozaki_energy_2000,castro_solution_2003,genovese_efficient_2006,wright_comparison_2006,ismail-beigi_truncation_2006,rozzi_exact_2006,genovese_efficient_2007,yu_equivalence_2008,dabo_electrostatics_2008,hine_electrostatic_2011}.
Among these approaches, the \textsc{BigDFT} Poisson solver~\cite{genovese_efficient_2006,genovese_efficient_2007} has proven successful in treating isolated- or surface-BC by solving the Hartree Potential in the correct BCs.

In this work, we also add the capability of treating reduced BCs to the plane-wave PAW implementation of \textsc{ABINIT}, by generalizing to PAW the \textsc{BigDFT} Poisson solver.
Here, all potentials and densities are calculated in real-space, as done within WVLs.
In particular, the local-ionic potential is calculated as explained in Section~\ref{sec.local-potential} and the Hartree potential is
obtained by using the \textsc{BigDFT} Poisson solver.
This approach can be accessed by setting the user-variable 
{\tt icoulomb} to  $1$ or $2$
for free- or surface- BCs, respectively.

\begin{figure}[h!]
\includegraphics{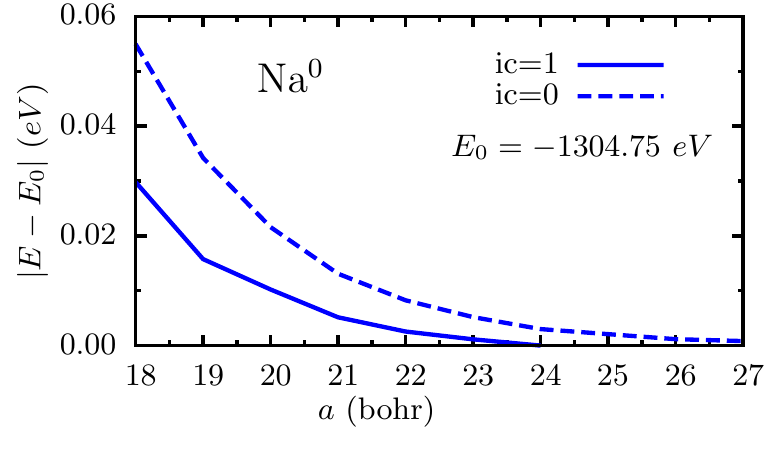}
\includegraphics{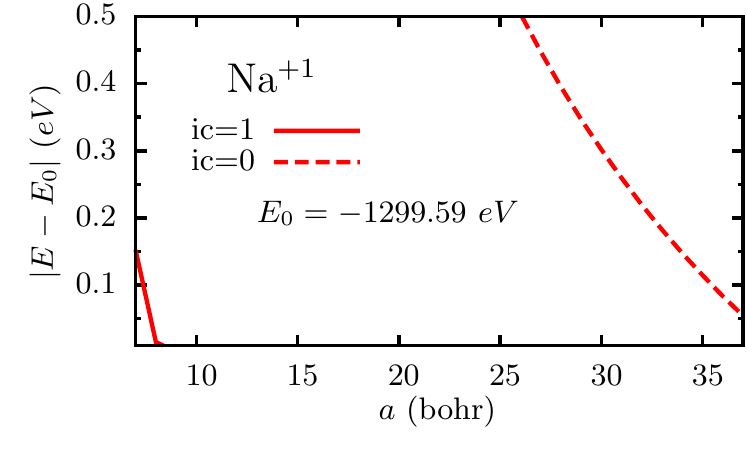}
\caption{(Color online)
Convergence on the lateral cell-size ($a$) for Na in 
a box.
The neutral (top) and charged (bottom) systems are considered.
The results with- (ic=1) and without- (ic=0)
the WVL Poisson solver are shown.
It is evident that using the 
WVL Poisson Solver (ic=0) 
drastically improves the convergence 
of charged species with respect
to the cell-size. 
}
\label{fig-psolver}
\end{figure}

To illustrate advantages of this implementation, we show in Figure~\ref{fig-psolver} the convergence with respect to the lateral unit-cell size ($a$) for Na in a squared-box.
The resulting total energies are $-$1304.75 and $-$1299.59~eV for the neutral and charged systems, respectively.
For the reciprocal space approach~(ic=0), these values are obtained after fitting the total energies to an infinite cell-size~($a\rightarrow \infty$), using the Makov-Payne method~\cite{makov.1995}.
As expected, convergence is achieved much faster using the WVL Poisson solver~(ic=1), since spurious interactions between neighboring cells are eliminated.

\bibliographystyle{elsarticle-num}
\bibliography{paper,PBC-electrostatics}







\end{document}